\documentclass[apj]{emulateapj}
\def\gsim{\ \raise 3pt \hbox{$\rangle$} \kern -8.5pt \raise -2pt \hbox{$\sim$}\ }
\usepackage{graphicx,natbib,color,amsmath,subfigure}
\begin{document}
\title{Cold, tenuous solar flare: acceleration without heating}
\author{Gregory D. Fleishman\altaffilmark{1,2}, Eduard P. Kontar\altaffilmark{3}, Gelu M. Nita\altaffilmark{1}, Dale E. Gary\altaffilmark{1}}
\altaffiltext{1}{Center For Solar-Terrestrial Research, New Jersey Institute of Technology, Newark, NJ 07102}
\altaffiltext{2}{Ioffe Physico-Technical Institute, St. Petersburg 194021, Russia}
\altaffiltext{3}{Department of Physics and Astronomy, University of
Glasgow, G12 8QQ, United Kingdom}

\begin{abstract}

We report the observation of an unusual cold, tenuous solar flare, which reveals itself via numerous
and prominent non-thermal manifestations, while lacking any noticeable thermal emission signature.
RHESSI hard X-rays and 0.1-18\hspace{0.1cm}GHz radio data from OVSA and Phoenix-2 show copious
electron acceleration  ($10^{35}$\hspace{0.1cm}electrons per second above 10\hspace{0.1cm}keV) typical for GOES M-class flares with electrons energies
up to $100$\hspace{0.1cm}keV, but GOES temperatures not exceeding $6.1$\hspace{0.1cm}MK. The imaging,
temporal, and spectral characteristics of the flare have led us
to a firm conclusion that the bulk of the microwave continuum emission from this flare
was produced directly in the acceleration region. The implications of this finding
for the flaring energy release and particle acceleration are discussed.

\end{abstract}


\section{Introduction}

An\hspace{0.1cm}outstanding question regarding solar flares is where, when, and how electrons
are  accelerated. The\hspace{0.1cm}direct detection of X-ray and radio emission from an acceleration region has
proved difficult. The\hspace{0.1cm}detection of X-rays from the\hspace{0.1cm}acceleration site
is challenging due to firstly a\hspace{0.1cm}relatively low density of the\hspace{0.1cm}surrounding
coronal plasma and secondly due to the\hspace{0.1cm}presence of competing emissions,
i.e., emission from hot flare loop plasma and trapped electron populations.
In addition, as the\hspace{0.1cm}hard X-ray (HXR)\hspace{0.1cm}flux is proportional to the\hspace{0.1cm}plasma density,
the\hspace{0.1cm}bulk of HXRs are emitted in the\hspace{0.1cm}dense plasma of the\hspace{0.1cm}chromosphere
(HXR footpoints) making X-ray imaging of tenuous coronal emission
problematic \citep{2003ApJ...595L..69L,emslie2003,2007LNP...725...65B}.
Studies of flares with footpoints occulted by the\hspace{0.1cm}solar disk \citep{krucker_lin_2008, krucker_etal_2010} provide direct imaging of the\hspace{0.1cm}looptop X-ray
emission, but are hampered because essential information on the\hspace{0.1cm}flare energy release contained in the\hspace{0.1cm}precipitating electrons becomes unavailable.  What is
needed is to cleanly separate the\hspace{0.1cm}acceleration and precipitation regions while retaining observations of both.  Having both radio and X-ray observations of
a flare without significant plasma heating and without noticeable magnetic trapping would provide the\hspace{0.1cm}needed information on both components to make
characterization of the\hspace{0.1cm}acceleration region possible.

Recently \citet{Bastian_etal_2007} have reported a\hspace{0.1cm}cold flare observed from a\hspace{0.1cm}very dense loop, where no significant heating occurred simply because the\hspace{0.1cm}flare energy was deposited to denser than usual plasma resulting in lower
than usual flaring plasma temperature. Although in their case highly important
implications for the\hspace{0.1cm}plasma heating and electron acceleration have been obtained,
the\hspace{0.1cm}strong Coulomb losses in the\hspace{0.1cm}dense coronal loop did not allow observing the\hspace{0.1cm}acceleration and thick-target regions separately.
From this perspective it would be better to have a\hspace{0.1cm}cold and \emph{tenuous}, rather than dense, flare.
However, such cold, tenuous flares seem to be unlikely because the\hspace{0.1cm}energy deposition from non-thermal
particles should result in even greater plasma heating in the\hspace{0.1cm}case of tenuous than dense plasma.
Nevertheless, inspection of available X-ray and radio databases reveals a\hspace{0.1cm}number of cold flare
candidates, one of the\hspace{0.1cm}most vivid examples of which is presented in this letter.

Specifically, we present and discuss an event (i)\hspace{0.1cm}lacking any noticeable GOES enhancement, (ii)\hspace{0.1cm}not showing any coronal X-ray source between the\hspace{0.1cm}footpoint sources, (iii)\hspace{0.1cm}with X-ray spectra
well fitted by a\hspace{0.1cm}thick-target model without any thermal component, and (iv)\hspace{0.1cm}producing relatively
low-frequency GS microwave continuum emission, which all together proves the\hspace{0.1cm}phenomenon
of the\hspace{0.1cm}cold, tenuous flare in the\hspace{0.1cm}case under study. The\hspace{0.1cm}available data offer evidence
that the\hspace{0.1cm}observed microwave GS emission is produced directly in the\hspace{0.1cm}acceleration
region of the\hspace{0.1cm}flare, and hence, parameters derived from microwave spectrum pertain to the\hspace{0.1cm}directly accelerated electron population.

\section{ X-ray and radio observations} 
\label{S_XR_observ}

\begin{figure}\centering
\includegraphics[width=0.9\columnwidth]{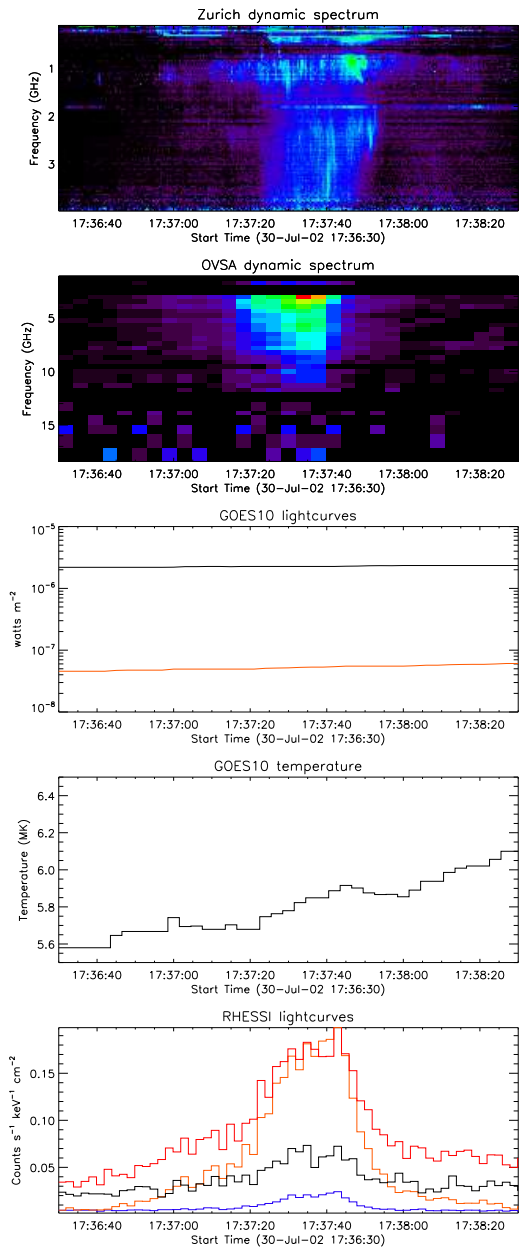}\\
\caption{\label{fig_30_jul_2002_over} Overview of  July 30, 2002 flare: Phoenix-2 and OVSA dynamic spectra, top panels. GOES (3\hspace{0.1cm}s) lightcurves and  temperature
as measured by GOES-10 spacecraft assuming photospheric abundances
from CHIANTI v5.2 (middle panels). RHESSI (2 second bins) lightcurves  (bottom panel)
in: 3-9\hspace{0.1cm}keV (black), 9-15\hspace{0.1cm}keV (red), 15-30\hspace{0.1cm}keV (orange), 30-100\hspace{0.1cm}keV (blue).  }
\end{figure}

\begin{figure}\centering
\includegraphics[width=0.9\columnwidth]{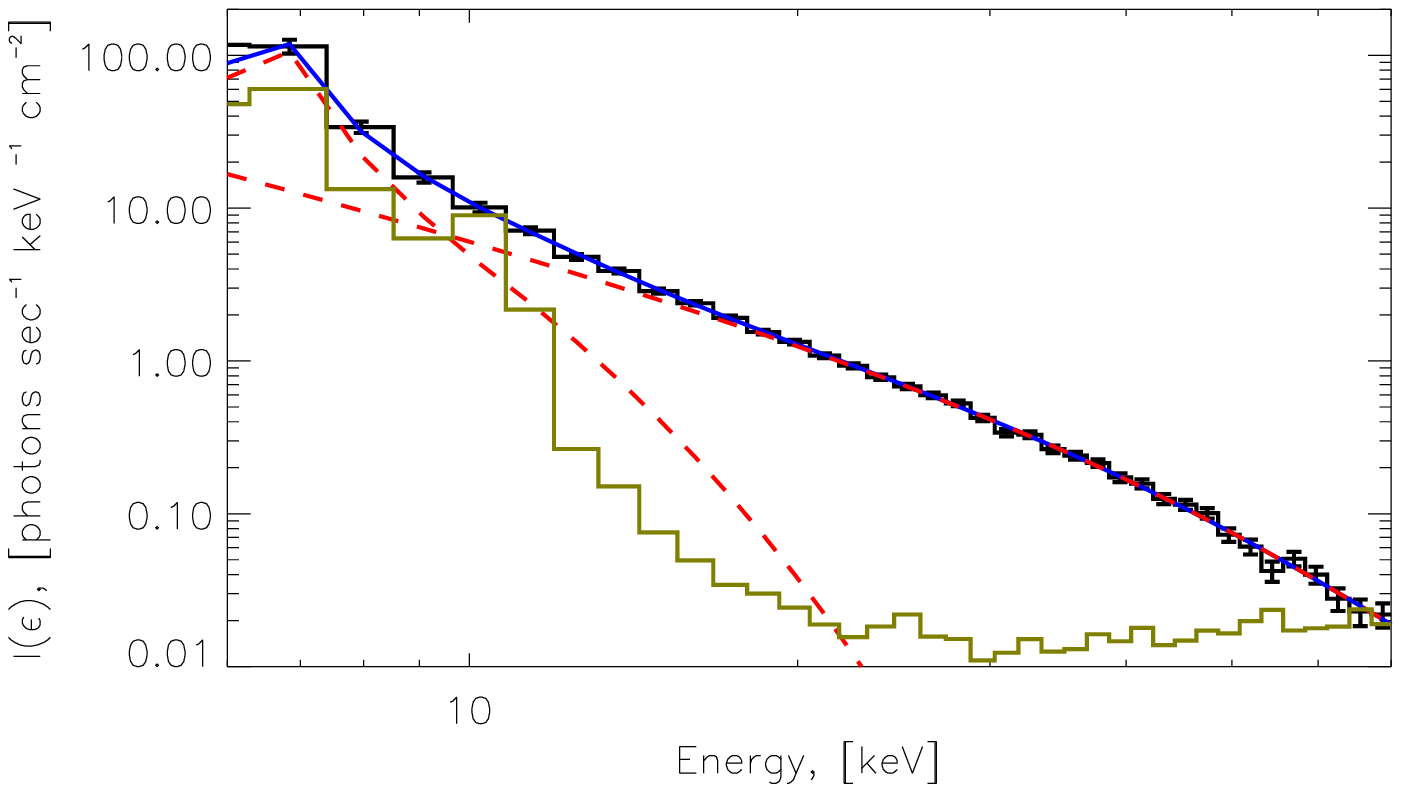} \\
\includegraphics[width=0.9\columnwidth]{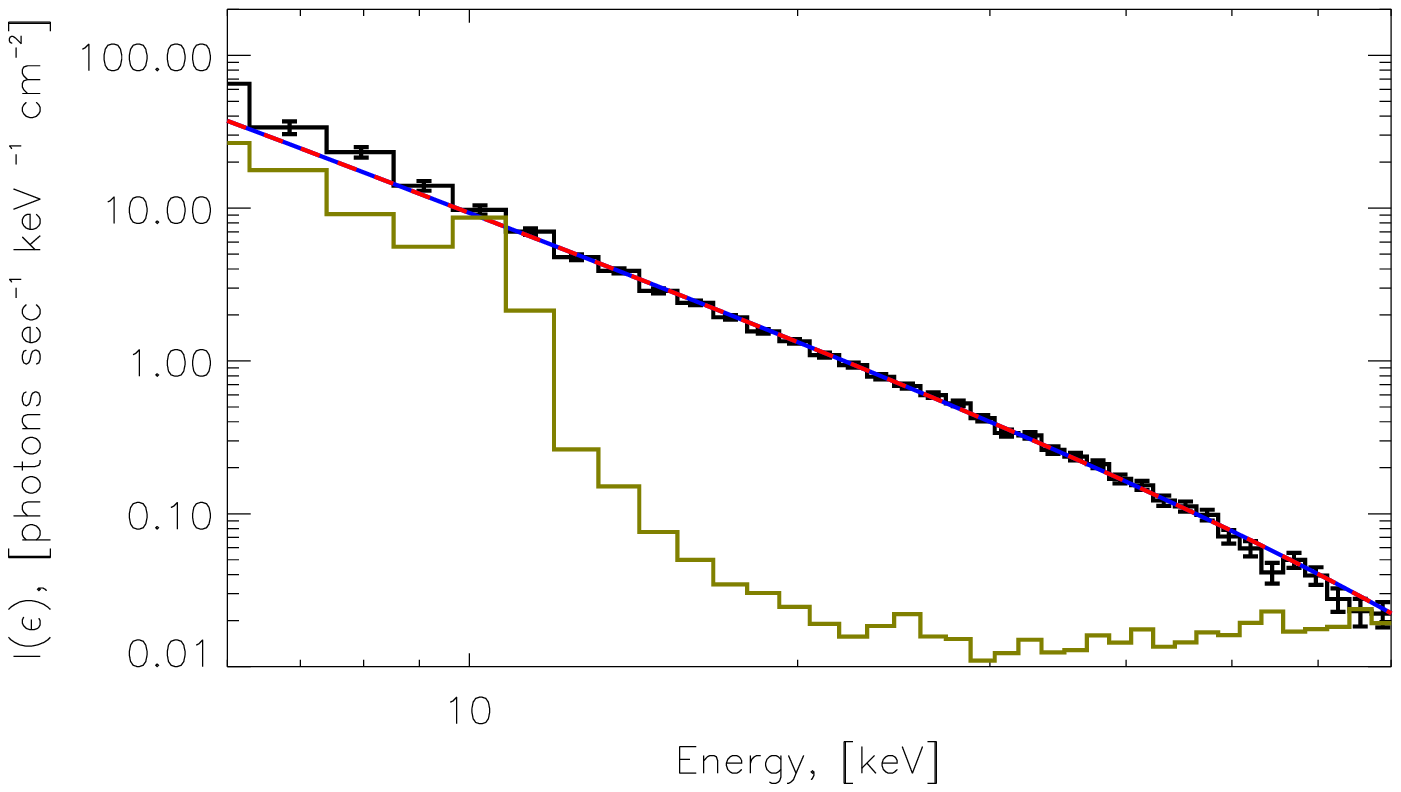}\\
\caption{\label{fig:peak_spectr} RHESSI X-ray spectrum (black) above 6\hspace{0.1cm}keV near the\hspace{0.1cm}peak of the\hspace{0.1cm}event
from $17:37:30-17:37:44$\hspace{0.1cm}UT using thermal (red-dashed) and thick-target model (red-dashed)
(top panel) with best fit (blue) parameters: emission measure $EM=9.7\times10^{45}$\hspace{0.1cm}cm$^{-3}$, temperature $16$\hspace{0.1cm}MK, $N_e(>6keV)=2.7\times10^{35}$\hspace{0.1cm}s$^{-1}$,
$\delta_{x}=3.56$. Thick-target model only (bottom panel): $N_e(>6keV)=2.8\times10^{35}$\hspace{0.1cm}s$^{-1}$, $\delta_{x}=3.56$. The pre-flare background is shown in brown. 
}
\end{figure}

\begin{figure}\centering
\qquad\includegraphics[width=0.9\columnwidth]{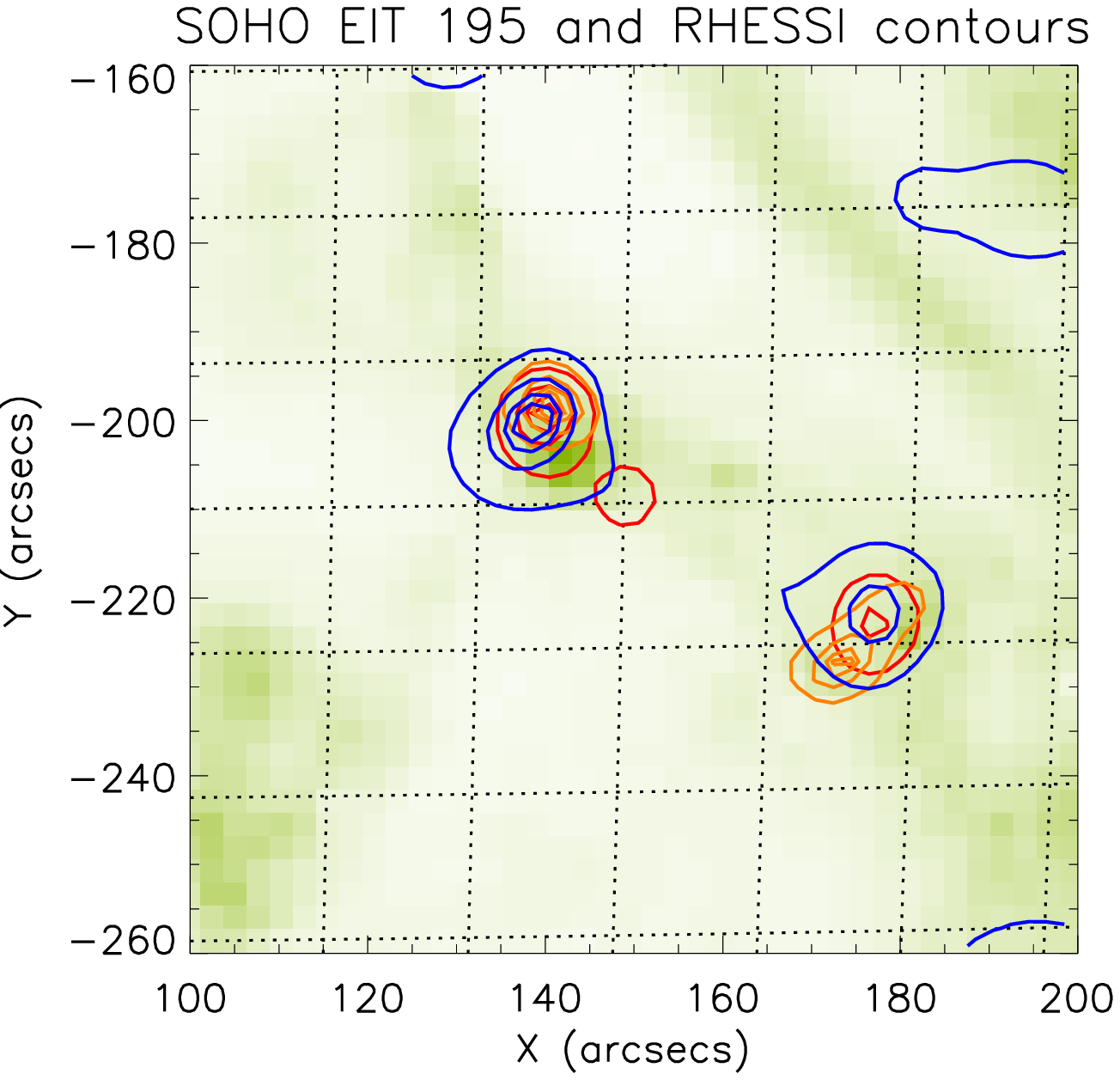} 
\includegraphics[width=0.8\columnwidth]{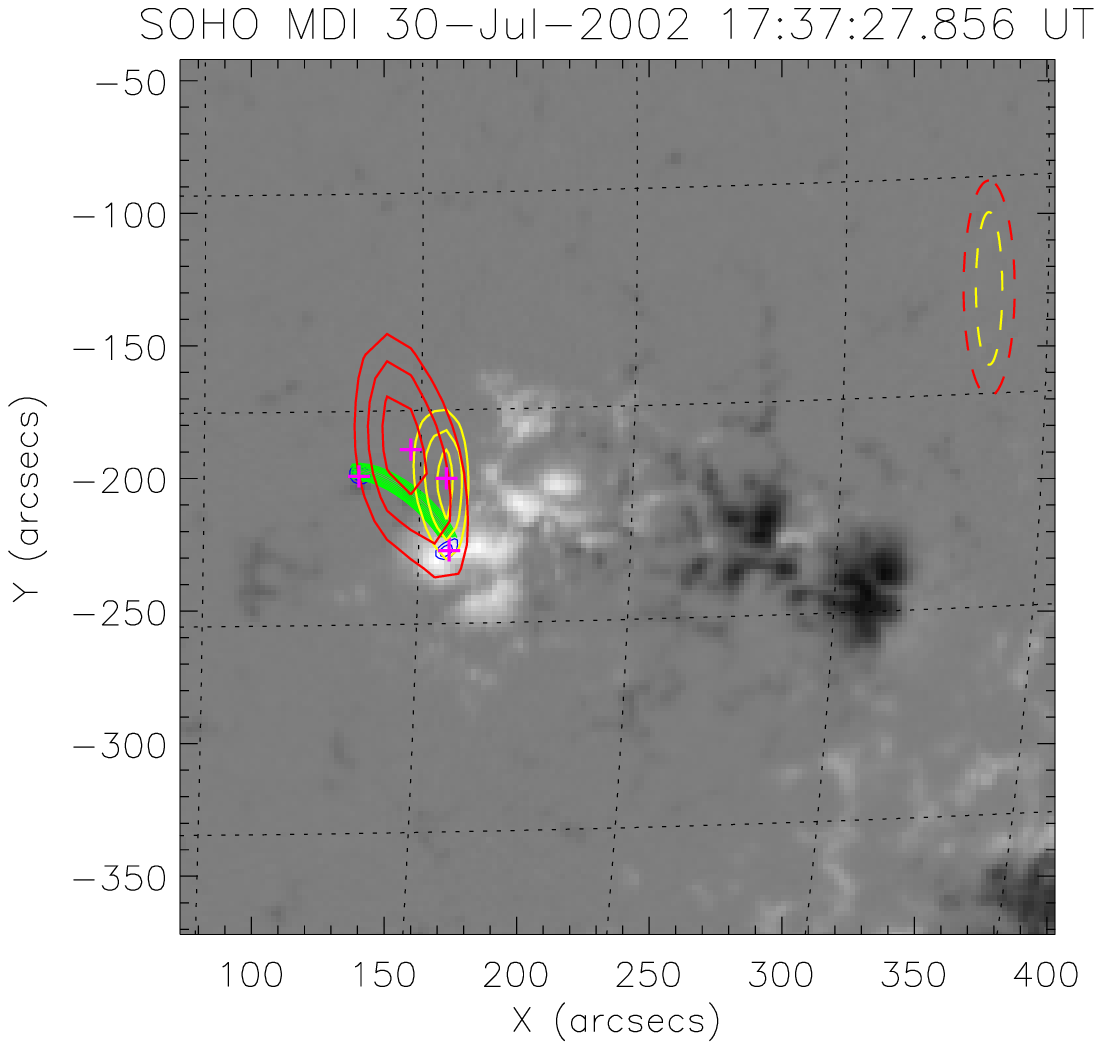}\\
\includegraphics[width=0.8\columnwidth]{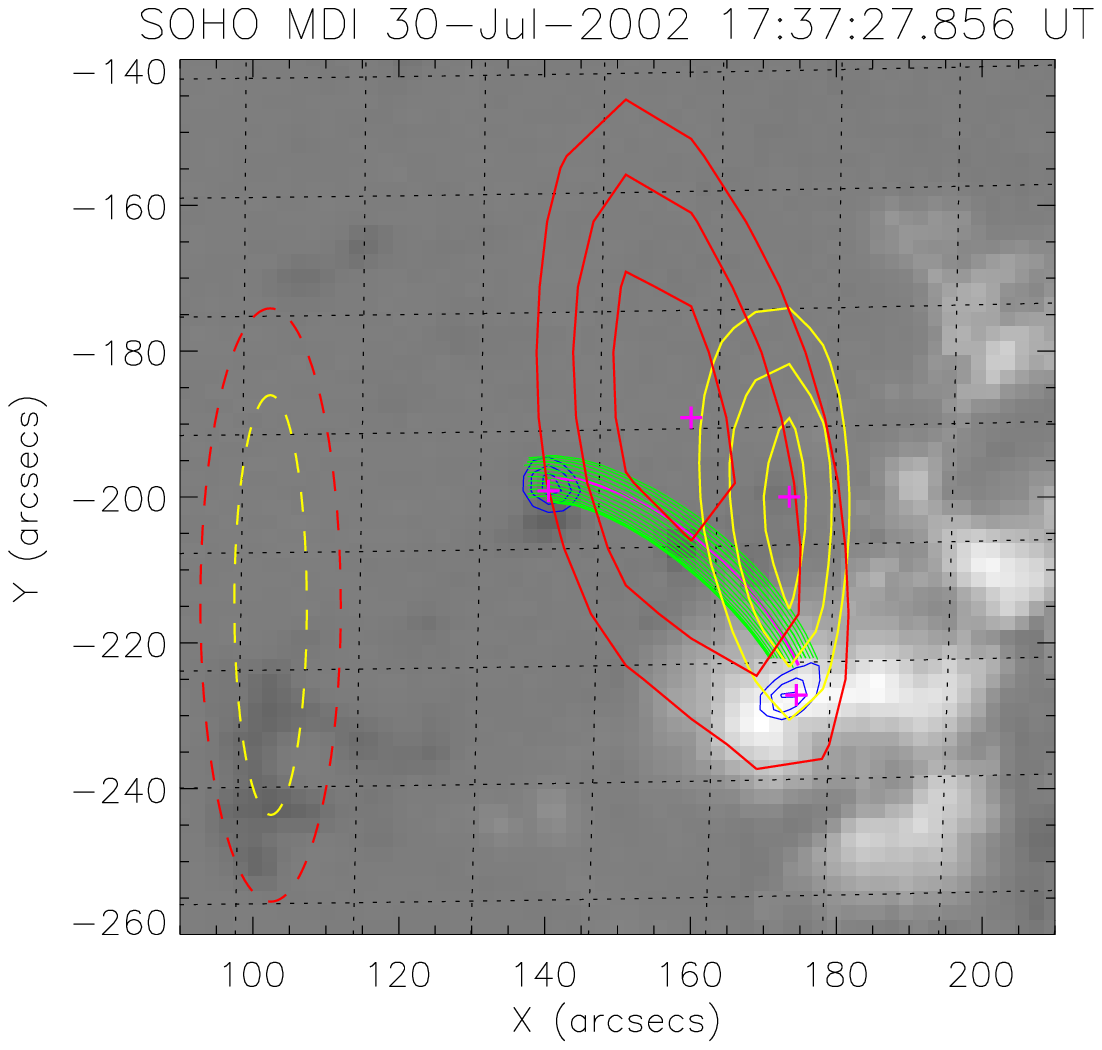}
\caption{\label{fig:Ximage} Top panel: Spatial distribution of X-ray emission from July 30, 2002 flare
in various energy ranges contours at 30, 50, 70, 90\% levels: $9-15$\hspace{0.1cm}keV (red) $15-30$\hspace{0.1cm}keV (orange), $30-100$\hspace{0.1cm}keV (blue).
Time accumulation interval for RHESSI images is $17:37-17:38$\hspace{0.1cm}UT. Background image is SoHO EIT\hspace{0.1cm}195
taken just before the\hspace{0.1cm}flare at $17:36$\hspace{0.1cm}UT. Middle and bottom panels: the\hspace{0.1cm}full and close-up view of the\hspace{0.1cm}active region and a\hspace{0.1cm}potential flux tube (green)
connecting two X-ray footpoints (blue contours),  $2.6-3.2$\hspace{0.1cm}GHz radio image (red contours)
and $4.2-8.2$\hspace{0.1cm}GHz (yellow contours). Magenta plus signs mark the\hspace{0.1cm}spatial peaks of the\hspace{0.1cm}HXR and radio sources. Dashed ellipses display the\hspace{0.1cm}sizes of the\hspace{0.1cm}synthesized beams.}
\end{figure}

The\hspace{0.1cm}July 30, 2002 flare appeared close to the\hspace{0.1cm}disk center (W10S07, NOAA AR10050)\hspace{0.1cm}and
is brightly visible in HXR images obtained by the\hspace{0.1cm}RHESSI spacecraft \citep{lin2002}.
Although the\hspace{0.1cm}radio emission recorded by Phoenix-2 spectrometer at 0.1--4\hspace{0.1cm}GHz \citep{1999SoPh..187..335M}
and Owens Valley Solar Array (OVSA) at 1--18\hspace{0.1cm}GHz also indicate an abundance of non-thermal
electrons, the\hspace{0.1cm}flare has  weak or nonexistent thermal soft X-ray emission (Figures\hspace{0.15cm}\ref{fig_30_jul_2002_over}--\ref{fig:Ximage}).
The\hspace{0.1cm}GOES light curves are almost flat at the\hspace{0.1cm}C2.2 level and the\hspace{0.1cm}temperature inferred
does not exceed $6.1$\hspace{0.1cm}MK. The\hspace{0.1cm}OVSA dynamic spectrum (4\hspace{0.1cm}s time resolution) displays
a low-frequency microwave continuum burst with a\hspace{0.1cm}peak frequency at 2--3\hspace{0.1cm}GHz.
The\hspace{0.1cm}Phoenix-2 dynamic spectrum obtained with higher (0.1\hspace{0.1cm}s) time resolution
shows that a\hspace{0.1cm}few type III-like features are superimposed on this
continuum at 1--4\hspace{0.1cm}GHz. There are also nonthermal emissions below 1\hspace{0.1cm}GHz.

The\hspace{0.1cm}spatially integrated RHESSI \citep{lin2002} X-ray spectrum
(Figure\hspace{0.15cm}\ref{fig:peak_spectr}) over the\hspace{0.1cm}duration of the\hspace{0.1cm}peak of the\hspace{0.1cm}flare indicates
strong non-thermal emission above $6$\hspace{0.1cm}keV and very weak or no thermal emission.
Spectral analysis (for spectroscopy, detectors 2 and 7 were avoided) was done using
OSPEX \citep{Schwartz_etal2002} with systematic errors
set to zero \citep[e.g.][]{Su_etal2009}. Since the\hspace{0.1cm}flare was close to the\hspace{0.1cm}disk center
(heliocentric angle $\sim20^o$), albedo correction was applied
\citep{kontar2006} assuming isotropic emission and hence minimum correction.
The\hspace{0.1cm}spectrum was fitted with the\hspace{0.1cm}standard thermal plus non-thermal thick-target
model \citep{brown1971} with $\chi^2{\sim}0.7$ (Figure\hspace{0.15cm}\ref{fig:peak_spectr}).
Bremsstrahlung cross-section following \citet{haug1997} has been utilized \citep{2006ApJ...643..523B}.
However, given the\hspace{0.1cm}clear lack of the\hspace{0.1cm}thermal emission attributes in the\hspace{0.1cm}event,
we also used a\hspace{0.1cm}purely non-thermal fit and found that the\hspace{0.1cm}count spectrum can
be nearly as well-fitted ($\delta_x{\sim}3.55$) without any thermal component,
$\chi^2\sim1.8$ (Figure\hspace{0.15cm}\ref{fig:peak_spectr}).

We have considered the\hspace{0.1cm}variation of spectral parameters over time using the\hspace{0.1cm}thick-target
model and fitting data above 10\hspace{0.1cm}keV (strong background below 10\hspace{0.1cm}keV does not allow reliable time-dependent fit at lower energies, Figure\hspace{0.15cm}\ref{fig:peak_spectr}). Both the\hspace{0.1cm}number
of accelerated electrons and the\hspace{0.1cm}spectral index demonstrate typical
soft-hard-soft behavior \citep[e.g.][]{1985SoPh..100..465D}. The\hspace{0.1cm}hardest
electron spectra $\delta_x{\sim}3.5$ are reached around
$17:37:40$\hspace{0.1cm}UT. At the\hspace{0.1cm}same time, the\hspace{0.1cm}electron acceleration rate
has its maximum  $F_{e\max}(>10keV){\simeq}10^{35}$\hspace{0.1cm}electrons
per second.

X-ray image reconstruction \citep{hurford2002} performed
with Pixon algorithm \citep{PinaPuetter1993} shows that the\hspace{0.1cm}flare
has two well defined footpoints (Figure\hspace{0.15cm}\ref{fig:Ximage}),
which are well visible over the\hspace{0.1cm}entire range of the\hspace{0.1cm}X-ray
spectrum $6-80$\hspace{0.1cm}keV. The\hspace{0.1cm}imaging below $10$\hspace{0.1cm}keV does not demonstrate
any thermal component in a\hspace{0.1cm}separate location as is often seen
at the\hspace{0.1cm}top of a\hspace{0.1cm}loop in flares \citep{Kosugi1992yohkoh,aschwanden2002,emslie2003,Kontar_etal08},
so all the\hspace{0.1cm}detectable X-ray emission down to the\hspace{0.1cm}lowest energy $\sim6$\hspace{0.1cm}keV  comes from the\hspace{0.1cm}footpoints. The\hspace{0.1cm}flare occurred
at the\hspace{0.1cm}extreme eastern edge of the\hspace{0.1cm}active region (Figure\hspace{0.15cm}\ref{fig:Ximage}) with the\hspace{0.1cm}weaker
X-ray source projected onto the\hspace{0.1cm}photosphere in a\hspace{0.1cm}region of strong positive magnetic field,
while the\hspace{0.1cm}stronger X-ray source projects onto a\hspace{0.1cm}small region of weaker
negative magnetic field, as has commonly been observed from asymmetric
flaring loops.

OVSA radio imaging for this flare is limited because only four (of six)
antennas (two big and two small antennas) recorded the\hspace{0.1cm}radio emission
at the\hspace{0.1cm}time of flare. To improve the\hspace{0.1cm}image quality a\hspace{0.1cm}frequency
synthesis in two separate bands,
$2.6-3.2$\hspace{0.1cm}GHz and $4.2-8.2$\hspace{0.1cm}GHz, with the\hspace{0.1cm}synthesized beam sizes of $19''\times81''$ and $10''\times58''$, respectively, was used. This suggests that the\hspace{0.1cm}both low- and high-frequency  radio sources are unresolved.
The\hspace{0.1cm}corresponding radio images (Figure\hspace{0.15cm}\ref{fig:Ximage})
are located between the\hspace{0.1cm}X-ray footpoints although with an offset
from their connecting line, which is consistent with the\hspace{0.1cm}radio sources placement
in a\hspace{0.1cm}coronal part of a\hspace{0.1cm}magnetic loop connecting the\hspace{0.1cm}X-ray footpoints.
The\hspace{0.1cm}higher-frequency radio source is displaced compared with the\hspace{0.1cm}lower-frequency
one towards the\hspace{0.1cm}stronger magnetic field (weaker X-ray) footpoint.
No spatial displacement with time is detected for either of the\hspace{0.1cm}radio sources.
Based on the\hspace{0.1cm}source separation, implied magnetic topology, and the\hspace{0.1cm}northern HXR source size, in
what follows  we adopt the\hspace{0.1cm}area of $10''$(transverse the\hspace{0.1cm}loop)$\times15''$(along the\hspace{0.1cm}loop),
and the\hspace{0.1cm}depth of $10''$ for the\hspace{0.1cm}lower-frequency radio source, which suggests the\hspace{0.1cm}radio source volume
of $V_{radio}\sim6\times10^{26}$\hspace{0.1cm}cm$^3$, and roughly half of that for transverse sizes
of the\hspace{0.1cm}higher-frequency source.

\section{Data Analysis}

Generally, GS continuum radio emission can be produced by any of (i)\hspace{0.1cm}a
magnetically trapped component \citep[trap-plus-precipitation model,][]{Melrose_Brown_1976},
or (ii)\hspace{0.1cm}a precipitating component, or (iii)\hspace{0.1cm}the\hspace{0.1cm}primary component within the\hspace{0.1cm}acceleration region.

These three populations of fast electrons produce radio emission with
distinctly different characteristics. Indeed, (i)\hspace{0.1cm}in the\hspace{0.1cm}case of magnetic trapping
the\hspace{0.1cm}electrons are accumulated at the\hspace{0.1cm}looptop \citep{melnikov_etal_2002}, and the\hspace{0.1cm}radio light curves are delayed
by roughly the\hspace{0.1cm}trapping time $\tau_{trap}$ relative to accelerator/X-ray light curves.
(ii)\hspace{0.1cm}In the\hspace{0.1cm}case of free electron propagation,
untrapped precipitating electrons are more evenly
distributed in a\hspace{0.1cm}tenuous loop, and no delay longer than $L/v$ is expected.
However, even with a\hspace{0.1cm}roughly uniform electron distribution, most of the\hspace{0.1cm}radio emission
comes from loop regions with the\hspace{0.1cm}strongest magnetic field.
Spectral indices of the\hspace{0.1cm}radio- and X-ray- producing fast electrons
differ by 1/2 from each other, because $L/v{\propto}E^{-1/2}$.
(iii)\hspace{0.1cm}In the\hspace{0.1cm}case of radio emission from the\hspace{0.1cm}acceleration region,
even though the\hspace{0.1cm}residence time ($\tau_l$)\hspace{0.1cm}that fast electrons spend in the\hspace{0.1cm}acceleration region can be relatively long,
the\hspace{0.1cm}radio and X-ray light curves are proportional
to each other simply because the\hspace{0.1cm}flux of the\hspace{0.1cm}X-ray producing electrons is equivalent to the\hspace{0.1cm}electron loss rate
from the\hspace{0.1cm}acceleration region, $F_e(t)=N_r(t)/\tau_l$.

\begin{figure}\centering
\includegraphics[width=0.9\columnwidth]{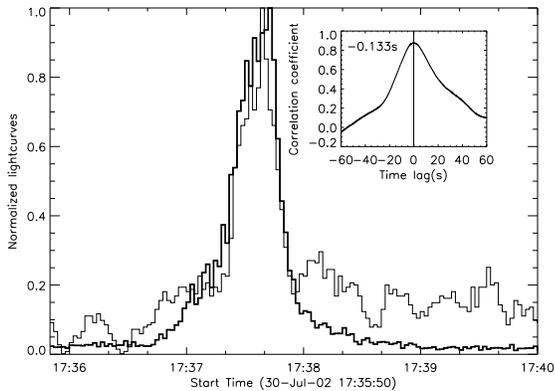}\\
\caption{\label{fig_r_z_corr} RHESSI 30-100\hspace{0.1cm}keV (thick line) and Phoenix-2 3.2 to 3.6\hspace{0.1cm}GHz (thin line) lightcurves of July 30, 2002 flare with 2\hspace{0.1cm}s time resolution both. The\hspace{0.1cm} light curves are highly correlated; no significant delay is present: the\hspace{0.1cm}lag correlation plot is given in the\hspace{0.1cm}insert; negative delay  means  the\hspace{0.1cm}radio emission comes first.}
\end{figure}

The\hspace{0.1cm}analysis of the\hspace{0.1cm}radio data requires, therefore, (in addition to the\hspace{0.1cm}electron injection
rate and spectrum derived from {\it RHESSI}) some information of the\hspace{0.1cm}fast electron residence
time at the\hspace{0.1cm}radio source. To address the\hspace{0.1cm}timing, we use Phoenix-2 (rather than the\hspace{0.1cm}OVSA)
data because of its higher time resolution. We select the\hspace{0.1cm}frequency range of 3.2 to 3.6\hspace{0.1cm}GHz
corresponding to the\hspace{0.1cm}optically thin part of the\hspace{0.1cm}radio spectrum and almost free of
fine structures and interference, see Figure\hspace{0.15cm}\ref{fig_30_jul_2002_over}.
The\hspace{0.1cm}cross-correlation (Figure\hspace{0.15cm}\ref{fig_r_z_corr}) displays clearly that the\hspace{0.1cm}radio
and HXR light curves are very similar to each other and there is no measurable
delay in the\hspace{0.1cm}radio component. In fact, the\hspace{0.1cm}cross-correlation is consistent with the\hspace{0.1cm}radio emission peaking $\sim130$\hspace{0.1cm}ms earlier. The\hspace{0.1cm}lack of noticeable delay between the\hspace{0.1cm}radio
and X-ray light curves is further confirmed by considering the\hspace{0.1cm}OVSA light
curves (4\hspace{0.1cm}s time resolution) at different optically thin frequencies.
Therefore, the\hspace{0.1cm}magnetically trapped electron component appears to be absent, 
and the\hspace{0.1cm}radio emission is formed by either  (ii)\hspace{0.1cm}precipitating electrons or
(iii)\hspace{0.1cm}electrons in the\hspace{0.1cm}acceleration region or both.

\begin{figure}\centering
\includegraphics[width=0.9\columnwidth]{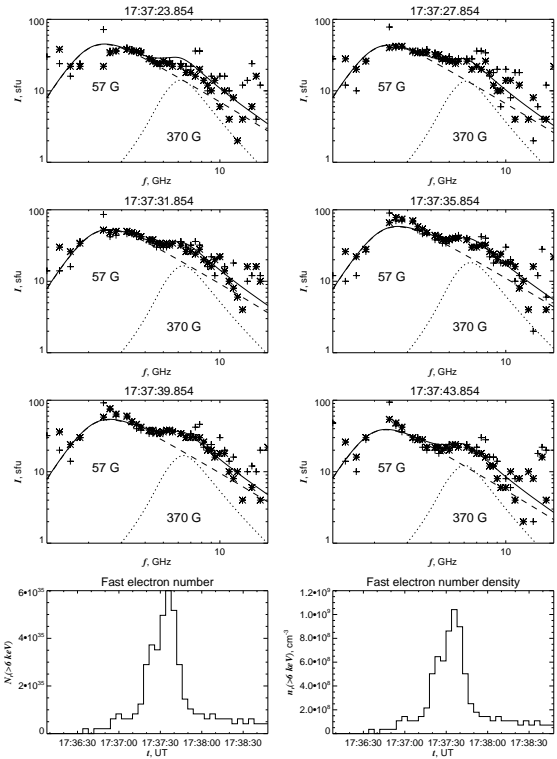}\\
\caption{\label{fig:fit_tt}
OVSA radio spectra obtained by two small antennas (pluses and asterisks; differences between them offer an idea about the\hspace{0.1cm}data scatter vs frequency) and model GS emission from the\hspace{0.1cm}acceleration region (dashed lines), precipitating electrons (dotted lines), and sum of these components (solid line). Total number and number density of the\hspace{0.1cm}fast electrons at the\hspace{0.1cm}radio source as derived from the\hspace{0.1cm}OVSA radio spectrum.}
\end{figure}

With the\hspace{0.1cm}spectrum of energetic electrons from HXR data, it is easy to estimate
the\hspace{0.1cm}radio emission produced by the\hspace{0.1cm}precipitating electron component.
Taking the\hspace{0.1cm}electron flux, the\hspace{0.1cm} spectral index of the\hspace{0.1cm}radio-producing
electrons $\delta_r{\approx}\delta_x+1/2$, and the\hspace{0.1cm}electron lifetime at the
loop $L/v$ (the\hspace{0.1cm}time of flight), we can vary the\hspace{0.1cm}magnetic field at the\hspace{0.1cm}source
in an attempt to match the\hspace{0.1cm}spectrum shape and flux level. However, if we match
the\hspace{0.1cm}spectrum peak position, we strongly underestimate the\hspace{0.1cm}radio flux, while if we match
the\hspace{0.1cm}flux level at the\hspace{0.1cm}peak frequency or at an optically thin frequency, we
overestimate the\hspace{0.1cm}spectrum peak frequency; examples of such spectra are given
in Figure\hspace{0.15cm}\ref{fig:fit_tt} by the\hspace{0.1cm}dotted curves. We conclude that precipitating
electrons only [option (ii)] cannot make a\hspace{0.1cm}dominant contribution to the\hspace{0.1cm}observed
radio spectrum.

To quantify the\hspace{0.1cm}third option, we have to consider further constraints
to estimate the\hspace{0.1cm}electron residence time in the\hspace{0.1cm}main radio source. On one hand, this residence
time must be shorter than the\hspace{0.1cm}radio light curve decay time, $\sim10$\hspace{0.1cm}s; otherwise, the\hspace{0.1cm}decay
of the\hspace{0.1cm}radio emission would be longer than observed. On the\hspace{0.1cm}other hand, the\hspace{0.1cm}extremely
low frequency of the\hspace{0.1cm}microwave spectrum peak implies the\hspace{0.1cm}magnetic field is well
below $100$\hspace{0.1cm}G at the\hspace{0.1cm}radio source, i.e., much smaller than the\hspace{0.1cm}footpoint magnetic field values
($\sim-130$\hspace{0.1cm}G and $\sim+1000$\hspace{0.1cm}G). This implies that the\hspace{0.1cm}residence time in the\hspace{0.1cm}main radio source
must be noticeably larger than the\hspace{0.1cm}time of flight, which is a\hspace{0.1cm}fraction
of second: otherwise, the\hspace{0.1cm}fast electron density would be evenly
distributed over the\hspace{0.1cm}loop and the\hspace{0.1cm}gyrosynchrotron (GS) microwave emission would
be dominantly produced at the\hspace{0.1cm}large field regions, resulting in a\hspace{0.1cm}spectrum
with much higher peak frequency than the\hspace{0.1cm}observed one (as in the\hspace{0.1cm}already considered
case of the\hspace{0.1cm}precipitating electron population). Thus, a\hspace{0.1cm}reasonable estimate
of this lifetime is somewhere between those two extremes, $\tau_{l}\sim 3$\hspace{0.1cm}s.

The\hspace{0.1cm}quality of the\hspace{0.1cm}OVSA data available for this event appears insufficient
to perform a\hspace{0.1cm}complete forward fit with all model parameters being free \citep{Fl_etal_2009},
which would require better calibrated imaging spectroscopy data.
Instead, we have to fix as many parameters as possible \citep{Bastian_etal_2007, Altyntsev_etal_2008}
and estimate one or two remaining parameters from the\hspace{0.1cm}fit. To do so we adopt the\hspace{0.1cm}maximum total
electron number $N_{r\max}(>6keV)=\tau_{l}F_{e\max}(>6keV)$ with a value of $F_{e\max}(>6keV)=2\times10^{35}$ electrons/s and $\tau_{l}=3$\hspace{0.1cm}s, and the\hspace{0.1cm}time evolution
of $N_{r}(>6keV)$ to be proportional to an optically thin gyrosynchrotron light curve.
The\hspace{0.1cm}total lifetime of the\hspace{0.1cm}electrons in the\hspace{0.1cm}flaring loop is a\hspace{0.1cm}sum of the\hspace{0.1cm}residence time
at the\hspace{0.1cm}radio source and the\hspace{0.1cm}time of flight between the\hspace{0.1cm}radio source and the\hspace{0.1cm}footpoints.
As we adopted a\hspace{0.1cm}constant, energy-independent, electron lifetime $\tau_{l}$ to be much larger
than the\hspace{0.1cm}time of flight ${\sim}L/v$ 
we have to accept for consistency that the\hspace{0.1cm}spectral index of the\hspace{0.1cm}radio emitting fast electrons
is roughly the\hspace{0.1cm}same as the\hspace{0.1cm}spectral index of HXR emitting fast electrons determined
above, $\delta_r=3.5$. The\hspace{0.1cm}radio source  sizes are taken as estimated in \S~\ref{S_XR_observ}.
The\hspace{0.1cm}thermal electron number density is adopted to be $n_{th}=1.5\times10^9$\hspace{0.1cm}cm$^{-3}$
(see the\hspace{0.1cm}next section); the\hspace{0.1cm}GS spectra
are not sensitive to this parameter until $n_{th}\lesssim2\times10^9$\hspace{0.1cm}cm$^{-3}$.
The\hspace{0.1cm}remaining radio source
parameter, not constrained by other observations, is the\hspace{0.1cm}magnetic field $B$, which is determined by
comparing the\hspace{0.1cm}observed (symbols) microwave and calculated (dashed curves)
GS spectra \citep{Fl_Kuzn_2010} in Figure\hspace{0.15cm}\ref{fig:fit_tt}. Remarkably, that the\hspace{0.1cm}whole time sequence of the\hspace{0.1cm}radio spectra
is reasonably fitted with a\hspace{0.1cm}single magnetic field strength of $B\approx60$\hspace{0.1cm}G; the\hspace{0.1cm}only source
parameter changing with time is the\hspace{0.1cm}instantaneous number of the\hspace{0.1cm}fast electrons, see Figure\hspace{0.15cm}\ref{fig:fit_tt}.
The\hspace{0.1cm}OVSA spectra, however, deviate from the\hspace{0.1cm}model dashed curves by the\hspace{0.1cm}presence of a\hspace{0.1cm}higher-frequency
bump at $f\sim4-8$\hspace{0.1cm}GHz. Nevertheless, adding the\hspace{0.1cm}contribution produced by precipitating
electrons (dotted curves) at a\hspace{0.1cm}larger magnetic field strength $B_{leg}$ peaking somewhere at the\hspace{0.1cm}western leg (60G$<B_{leg}<$1000G) of the\hspace{0.1cm} loop (perhaps, around the\hspace{0.1cm}mirror point) as follows from the\hspace{0.1cm}OVSA imaging, Figure\hspace{0.15cm}\ref{fig:Ximage}, offers a\hspace{0.1cm}nice, consistent overall fit (solid curves) to the\hspace{0.1cm}spectra.
We conclude that the\hspace{0.1cm}radio spectrum is dominated by the\hspace{0.1cm}GS emission
from the\hspace{0.1cm}electron acceleration region with a\hspace{0.1cm}distinct weaker contribution
from the\hspace{0.1cm}precipitating electrons.

\section{Discussion}

In order to complete a\hspace{0.1cm}model for this cold flare event we have to estimate the\hspace{0.1cm}flaring
loop geometry. As a\hspace{0.1cm}zero-order approximation we utilize a\hspace{0.1cm}potential field extrapolation \citep{Rudenko_2001}
based on the\hspace{0.1cm}SoHO/MDI line-of-sight magnetogram. Figure\hspace{0.15cm}\ref{fig:Ximage} shows
that there is a\hspace{0.1cm}flux tube connecting the\hspace{0.1cm}two X-ray footpoints, which confirms the
existence of the\hspace{0.1cm}required magnetic connectivity. This magnetic loop is highly asymmetric
with the\hspace{0.1cm}magnetic field reaching its minimum value (around 130\hspace{0.1cm}G) at the\hspace{0.1cm}northern
footpoint (stronger X-ray source). The\hspace{0.1cm}length of the\hspace{0.1cm}central field line is
about $55''\approx4\times10^9$\hspace{0.1cm}cm. We know, however, that the\hspace{0.1cm}flare
phenomenon requires a\hspace{0.1cm}non-potential magnetic loop. Furthermore, the\hspace{0.1cm}magnetic field at the\hspace{0.1cm}radio
source (which is likely to belong to the\hspace{0.1cm}same magnetic structure because of excellent
light curve correlation) is below 100\hspace{0.1cm}G, which is likely to be located higher
than the\hspace{0.1cm}potential loop in Figure\hspace{0.15cm}\ref{fig:Ximage}. We, therefore, propose that
the\hspace{0.1cm}true flaring loop is higher and the\hspace{0.1cm}length of the\hspace{0.1cm}central field line
is somewhat longer; for the\hspace{0.1cm}estimate we adopt $L\sim100''\sim7\times10^9$\hspace{0.1cm}cm that
yields the\hspace{0.1cm}loop volume $V_{loop}\sim4\times10^{27}$\hspace{0.1cm}cm$^3$ roughly 5 times larger
than the\hspace{0.1cm}radio source volume.

The\hspace{0.1cm}next needed step is an estimate of the\hspace{0.1cm}thermal density in the\hspace{0.1cm}flaring loop.
From the\hspace{0.1cm}radio spectra and from the\hspace{0.1cm}absence of any coronal X-ray source we already know that this density is low.  In fact, both radio and X-ray data can
be fit  without any thermal plasma at all.
Let us consider the\hspace{0.1cm}10keV electron Coulomb losses
to quantify the\hspace{0.1cm}thermal density. The\hspace{0.1cm}collisional lifetime of the\hspace{0.1cm}fast
electrons is $t_E\approx20E_{100}^{3/2}n_{10}^{-1}$\hspace{0.1cm}s \citep{Bastian_etal_2007}, which
yields $t_{10keV}\lesssim3$\hspace{0.1cm}s for the\hspace{0.1cm}background plasma density $n_{th}\gtrsim 2\times10^{9}$\hspace{0.1cm}cm$^{-3}$,
which would imply the\hspace{0.1cm}presence of a\hspace{0.1cm}coronal 10keV X-ray component in contradiction
with observations. We conclude that $2\times10^{9}$\hspace{0.1cm}cm$^{-3}$ is an upper bound
for the\hspace{0.1cm}thermal electron density. To estimate the\hspace{0.1cm}lower bound of the\hspace{0.1cm}density, we consider
the\hspace{0.1cm}efficiency of the\hspace{0.1cm}electron accelerator. From the\hspace{0.1cm}X-ray fit we know that the\hspace{0.1cm}peak acceleration
efficiency is at least $F_{e\max}(>6keV)\sim 2{\times}10^{35}$ electrons per second.
The\hspace{0.1cm}duration of the\hspace{0.1cm}flare at the\hspace{0.1cm}half of the\hspace{0.1cm}peak level is about $t_{1/2}\approx25$\hspace{0.1cm}s,
thus the\hspace{0.1cm}total number of accelerated electrons is $N_{tot}{\sim}t_{1/2}F_{e\max}\sim\hspace{0.1cm}5\times10^{36}$.
These electrons must apparently be taken from the\hspace{0.1cm}thermal electrons available in the\hspace{0.1cm}flaring
loop prior to the\hspace{0.1cm}flare, thus, the\hspace{0.1cm}ratio of $N_{tot}/V_{loop}{\sim}10^9$\hspace{0.1cm}cm$^{-3}$
represents a\hspace{0.1cm}lower bound of the\hspace{0.1cm}thermal electron density. These estimates justify
the\hspace{0.1cm}$n_{th}$ value adopted in the\hspace{0.1cm}previous section.

Let us proceed now to the\hspace{0.1cm}energy release and plasma heating efficiency. The\hspace{0.1cm}energy release
rate $dW/dt$ is estimated as the\hspace{0.1cm}product of the\hspace{0.1cm}minimum energy (6keV) and the\hspace{0.1cm}acceleration
rate $F_{e}(>6keV)$, which yields $\sim 2\times10^{27}$\hspace{0.1cm}ergs/s at the\hspace{0.1cm}flare peak time. Being evenly
distributed over the\hspace{0.1cm}loop volume this corresponds to the\hspace{0.1cm}averaged density of the\hspace{0.1cm}energy release of $\sim0.5$\hspace{0.1cm}erg/cm$^{-3}$/s and, being multiplied by the\hspace{0.1cm}effective duration $t_{1/2}$, the\hspace{0.1cm}energy
density deposition of $w\sim12$\hspace{0.1cm}erg/cm$^{-3}$. Most of this energy is produced in the\hspace{0.1cm}form
of accelerated electrons around 10\hspace{0.1cm}keV. During the\hspace{0.1cm}time of flight in the\hspace{0.1cm}loop
(with density $1.5\times10^{9}$\hspace{0.1cm}cm$^{-3}$ and half length $\sim3\times10^9$\hspace{0.1cm}cm) these electrons
lose about $\Delta\simeq15\%$ of their initial energy. Thus, we can estimate the\hspace{0.1cm}plasma heating
by the\hspace{0.1cm}accelerated electrons up to $T{\simeq}w/(k_Bn_{th})\times\Delta\sim7$\hspace{0.1cm}MK,
where $k_B$ is the\hspace{0.1cm}Boltzman constant. Combined with a\hspace{0.1cm}relatively low emission measure
of the\hspace{0.1cm}tenuous loop, this heating is undetectable by GOES and RHESSI, even though the\hspace{0.1cm}acceleration
efficiency is extremely high. 

Many acceleration mechanisms require the\hspace{0.1cm}particles to be confined
in the\hspace{0.1cm}acceleration region.
In particular, this is the\hspace{0.1cm}case for acceleration in  collapsing
magnetic traps \citep[e.g.][]{1975ApJ...200..734B,1997ApJ...485..859S,2004A&A...419.1159K}
and for stochastic acceleration \citep[e.g.][]{1993ApJ...418..912L,1996ApJ...461..445M,Byk_Fl_2009,2010A&A...519A.114B}, but not for a\hspace{0.1cm}DC field acceleration.
This flare does not display
any characteristics expected for the\hspace{0.1cm}collapsing trap scenario
\citep[e.g., a spatial displacement of the\hspace{0.1cm}source, the\hspace{0.1cm}magnetic field growth, the\hspace{0.1cm}radio peak frequency increase, or a time delay between radio light curves at different frequencies;][]{Li_Fl_2009},
while it is consistent with the\hspace{0.1cm}stochastic acceleration \citep{Park_Fl_2010} in a\hspace{0.1cm}magnetic loop, when a standard, relatively narrowband, GS emission is produced at a given volume (permitted with a loop magnetic field) by the\hspace{0.1cm}electrons accelerated there by a turbulence, whose side effect is to enhance the\hspace{0.1cm}electron trapping and so increase, as observed, their residence time at the\hspace{0.1cm} acceleration region.
We conclude that detection of GS radio emission from a\hspace{0.1cm}region of stochastic
acceleration of electrons is likely in the\hspace{0.1cm}event
under study. Various stochastic acceleration scenarios differ from each other in some predictions,
and so in principle are distinguishable by observations. One of them is the\hspace{0.1cm}energy dependence of the\hspace{0.1cm}fast
electron lifetime at the\hspace{0.1cm}acceleration region. Our data are consistent with $\tau_l(E)=const$,
which is the\hspace{0.1cm}case, in particular, for acceleration by  strong turbulence \citep{Byk_Fl_2009}.
However, the\hspace{0.1cm}precision of this constancy is insufficient in our observations to exclude competing
versions of the\hspace{0.1cm}stochastic acceleration.

What is special about this flare, which allowed detecting the\hspace{0.1cm}radio emission
from the\hspace{0.1cm}acceleration region, compared with other flares where the\hspace{0.1cm}radio emission is
often dominated by trapped population of fast electrons?
Possibly, an important aspect is the\hspace{0.1cm}exceptional asymmetry of the\hspace{0.1cm}flare loop; recall that the\hspace{0.1cm}corresponding
potential loop displays the\hspace{0.1cm}minimum strength of the\hspace{0.1cm}magnetic field absolute value at the\hspace{0.1cm}northern
photospheric footpoint (rather than coronal level), making the\hspace{0.1cm}magnetic trapping of fast electrons
totally inefficient in this case.
A\hspace{0.1cm}further important point is the\hspace{0.1cm}somewhat low initial plasma density of the\hspace{0.1cm}loop. This low density
maintains relatively low coronal Coulomb losses of the\hspace{0.1cm}accelerated electrons, which allows
them to freely stream towards the\hspace{0.1cm}chromosphere to produce the\hspace{0.1cm}X-ray emission. In addition,
the\hspace{0.1cm}low density cannot supply the\hspace{0.1cm}accelerator with seed electrons longer than half a\hspace{0.1cm}minute
or so, thus, the\hspace{0.1cm}accelerator becomes exhausted very quickly, before the\hspace{0.1cm}energy needed
for substantial (measurable) plasma heating has been released. This explains why the\hspace{0.1cm}main
plasma remains relatively cold although the\hspace{0.1cm}acceleration efficiency is exceptionally
high (almost all available electrons are being eventually accelerated).

\section{Conclusions}

We have reported the\hspace{0.1cm}observations and analysis of a\hspace{0.1cm}cold, tenuous flare, which displays prominent and numerous non-thermal signatures,
while not showing any evidence for thermal plasma heating or chromospheric evaporation.
A\hspace{0.1cm}highly asymmetric flaring loop, combined with rather low thermal electron density, made
it possible to detect the\hspace{0.1cm}GS radio emission directly from the\hspace{0.1cm}acceleration site.
We found that the\hspace{0.1cm}electron acceleration efficiency is very high in the\hspace{0.1cm}flare,
so almost all available thermal electrons are eventually accelerated. Some sort of stochastic acceleration process is needed to account for an approximately energy-independent lifetime of about 3\hspace{0.1cm}s for the\hspace{0.1cm}electrons in the\hspace{0.1cm}acceleration region.
We emphasize that the\hspace{0.1cm}numbers derived for the\hspace{0.1cm}quantity of accelerated electrons
offer a\hspace{0.1cm}lower bound for this measure because for our estimates we adopted a\hspace{0.1cm}lower bound of the\hspace{0.1cm}electron
flux ($F_{e\max}(>6keV)\simeq2\times10^{35}$\hspace{0.1cm}electrons
per second) and lowest electron spectral index ($\delta_r=3.5$) compatible with the\hspace{0.1cm}HXR fit;
taking mean values would result in even more powerful acceleration. However, given a\hspace{0.1cm}relatively
small flaring volume and rather low thermal density at the\hspace{0.1cm}flaring loop, the\hspace{0.1cm}total energy
release turned out to be insufficient for a\hspace{0.1cm}significant heating of the\hspace{0.1cm}coronal
plasma or for a\hspace{0.1cm}prominent chromospheric response giving rise to chromospheric
evaporation.

\acknowledgments
This work was supported in part by NSF grants
AGS-0961867, AST-0908344, and NASA grants NNX10AF27G and NNX11AB49G to New Jersey
Institute of Technology,
and by the\hspace{0.1cm}RFBR  grants   09-02-00226 and 09-02-00624.
This work was supported by a\hspace{0.1cm}UK STFC rolling grant, STFC/PPARC Advanced
Fellowship, and the\hspace{0.1cm}Leverhulme Trust, UK. Financial support by the
European Commission through the\hspace{0.1cm}SOLAIRE and HESPE Networks is
gratefully acknowledged. The\hspace{0.1cm}authors are greatly thankful to George Rudenko and Vasil Yurchishin for valuable discussions of the\hspace{0.1cm}coronal magnetic extrapolations.

\end{document}